\documentclass[conference]{IEEEtran}
\IEEEoverridecommandlockouts

\usepackage{cite}
\usepackage{amsmath,amssymb,amsfonts}
\usepackage{algorithmic}
\usepackage{graphicx}
\usepackage{textcomp}
\usepackage{xcolor}
\usepackage{comment}
\usepackage[acronym]{glossaries}
\usepackage[free-standing-units,per-mode=repeated-symbol,binary-units=true,detect-weight=true,detect-family=true]{siunitx}[=v2]
\usepackage{lscape}
\usepackage{multirow}

\usepackage{threeparttable} % For notes under the table
\usepackage{booktabs}  
\usepackage{tikz}
\newcommand*\circled[1]{\tikz[baseline=(char.base)]{
            \node[shape=circle,draw,inner sep=1pt] (char) {#1};}}
\def\BibTeX{{\rm B\kern-.05em{\sc i\kern-.025em b}\kern-.08em
    T\kern-.1667em\lower.7ex\hbox{E}\kern-.125emX}}

\begin{document}

\title{A Dynamic Allocation Scheme for Adaptive Shared-Memory Mapping on Kilo-core RV Clusters for Attention-Based Model Deployment\\
}

\ifx\blind\undefined
    \author{
    \IEEEauthorblockN{Bowen Wang}
    \IEEEauthorblockA{
    ETH Z\"{u}rich\\
    Z\"{u}rich, Switzerland\\
    bowwang@iis.ee.ethz.ch} \\
    \and
    \IEEEauthorblockN{Marco Bertuletti}
    \IEEEauthorblockA{
    ETH Z\"{u}rich\\
    Z\"{u}rich, Switzerland\\
    mbertuletti@iis.ee.ethz.ch} \\
    \and
    \IEEEauthorblockN{Yichao Zhang}
    \IEEEauthorblockA{
    ETH Z\"{u}rich\\
    Z\"{u}rich, Switzerland\\
    yiczhang@iis.ee.ethz.ch}\\
    
    \and
    \IEEEauthorblockN{Victor J.B. Jung}
    \IEEEauthorblockA{
    ETH Z\"{u}rich\\
    Z\"{u}rich, Switzerland\\
    jungvi@iis.ee.ethz.ch}\\
    \and
    \IEEEauthorblockN{Luca Benini}
    \IEEEauthorblockA{ETH Z\"{u}rich\\
    Z\"{u}rich, Switzerland\\
    Universit\`a di Bologna \\
    Bologna, Italy \\
    lbenini@iis.ee.ethz.ch}
    }
\else
    \author{\centering{\textit{Authors omitted for blind review.}\vspace{2cm}}}
\fi
\newacronym{LLM}{LLM}{Large Language Model}
\newacronym{ai}{AI}{Artificial Intelligence}
\newacronym{dl}{DL}{Deep Learning}
\newacronym{ml}{ML}{Machine Learning}
\newacronym{ViT}{ViT}{Vision Transformer}
\newacronym{DAS}{DAS}{Dynamic Allocation Scheme}
\newacronym{NoC}{NoC}{Networks-on-Chip}
\newacronym{PE}{PE}{processing element}
\newacronym{pe}{PE}{processing element}
\newacronym{SM}{SM}{Streaming Multiprocessor}
\newacronym{TSP}{TSP}{Tensor Streaming Processor}
\newacronym{GPU}{GPU}{Graphics Processing Unit}
\newacronym{DMA}{DMA}{Direct Memory Access}
\newacronym{spm}{SPM}{Scratchpad Memory}
\newacronym{NUMA}{NUMA}{Non-Uniform Memory Access}
\newacronym{LSU}{LSU}{Load-Store Unit}
\newacronym{csr}{CSR}{Control/State Register}
\newacronym{API}{API}{Application Programming Interface}
\newacronym{MAC}{MAC}{Multiply-Accumulate}

% Kernel Related

\newacronym{AXPY}{AXPY}{AX-Plus Y}
\newacronym{GEMV}{GEMV}{General Matrix-Vector Multiplication}
\newacronym{GEMM}{GEMM}{General Matrix Multiplication}
\newacronym{FA2}{FA2}{Flash-Attention 2}
\newacronym{MHA}{MHA}{Multi-Head Attention}
\newacronym{MHSA}{MHSA}{Multi-Head Self-Attention}
\newacronym{vit}{ViT}{Vision Transformer}
\newacronym{FF}{FF}{Feed-Forward}
\newacronym{dotp}{DotP}{Dot-Product}

\newacronym{CG}{CG}{Computation Group}
\newacronym{IPC}{IPC}{Instruction per Cycle}
\newacronym{ipc}{IPC}{Instruction per Cycle}
\newacronym{qkv gen}{QKV Gen}{Query-Key-Value Generation}
\newacronym{sa}{SA}{Self-Attention}
\newacronym{op}{OP}{Output Projection}
\newacronym{flop}{FLOP}{Floating-Point Operation}

\newacronym{spmd}{SPMD}{Single-Program Multiple-Data}
\newacronym{SPMD}{SPMD}{Single-Program Multiple-Data}
\newacronym{SIMT}{SIMT}{Single-Instruction Multiple-Thread}
\newacronym{pim}{PIM}{Processing-in-Memory}
\newacronym{PIM}{PIM}{Processing-in-Memory}
\maketitle

\begin{abstract}
Attention-based models demand flexible hardware to manage diverse kernels with varying arithmetic intensities and memory access patterns. Large clusters with shared L1 memory, a common architectural pattern, struggle to fully utilize their \glspl{PE} when scaled up due to reduced throughput in the hierarchical \gls{PE}-to-L1 intra-cluster interconnect. This paper presents \gls{DAS}, a runtime programmable address remapping hardware unit coupled with a unified memory allocator, designed to minimize data access contention of \glspl{PE} onto the multi-banked L1. We evaluated \gls{DAS} on an aggressively scaled-up 1024-\gls{PE} RISC-V cluster with \gls{NUMA} \gls{PE}-to-L1 interconnect to demonstrate its potential for improving data locality in large parallel machine learning workloads. For a \gls{ViT}-L/16 model, each encoder layer executes in \SI{5.67}{\milli\second}, achieving a 1.94$\times$ speedup over the fixed word-level interleaved baseline with 0.81 PE utilization. Implemented in 12nm FinFET technology, \gls{DAS} incurs \textless \SI{0.1}{\percent} area overhead.

\end{abstract}


\glsresetall
\section{Introduction}
\begin{figure}[ht]
  \centering
  \includegraphics[width=\linewidth]{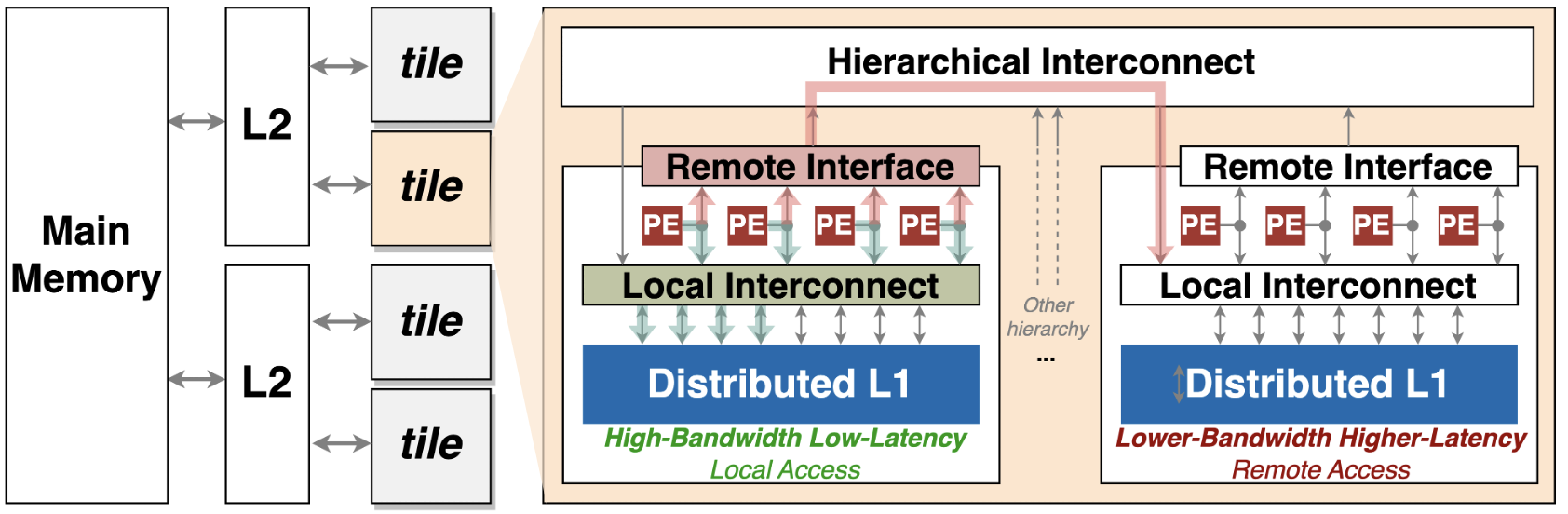}
  \caption{Many-core cluster architecture. Tiles are connected to a hierarchical memory system via high-throughput interconnects. Within each tile, \glspl{PE} share an L1 memory through a low-latency, high-bandwidth hierarchical in-tile interconnect.}
  \label{fig:generic_cluster}
    \vspace{-1em}
\end{figure}

The rapid growth of \gls{ai} workloads accelerates the development of processing platforms with increasing numbers of \glspl{PE} tightly integrated with on-chip memory. As depicted in Fig.\ref{fig:generic_cluster}, these platforms—whether implemented on single-chip designs\cite{MAIA}, chiplets\cite{DWAVE}, or wafer-scale engines~\cite{CEREBRAS}—group \glspl{PE} in \textit{tiles}, with high-throughput interconnects to a hierarchical memory system. A tile forms a tightly coupled many-core cluster, characterized by low-latency access of \glspl{PE} to local, in-tile memory (hereafter referred to as L1 memory).

In the last few years, Transformer-based models for different applications expanded in size\cite{model_scaling}, with rapidly growing larger attention blocks and longer sequence lengths: in language processing \cite{llama2}, computer vision \cite{vit_2020}, and time-series analysis\cite{time_series}, reaching hundred billions of parameters. From the hardware side, this complexity scaling is harnessed by scaling out (i.e., increasing the number of tiles) and scaling up (i.e., increasing the PE count and memory capacity of the tile).  

In this work, we focus on the scale-up dimension, ensuring that larger clusters remain efficient, achieving high PE utilization. This goal presents significant challenges, especially in the context of transformer inference, where computational patterns vary widely\cite{llm_inference}, often involving an alternation between \gls{GEMM} and \gls{GEMV}, with additional stages such as Add-and-Norm layers handling residual connections and layer normalization. 

State-of-the-art hardware platforms adopt distinct strategies to achieve efficient execution for their compute tiles on heterogeneous workloads. NVIDIA’s \gls{GPU} tiles, namely the \gls{SM}\cite{nvidia_h100}, leverage a fine-grained \gls{SIMT} programming model to concurrently manage 2048 threads for latency tolerance. Cerebras’s \mbox{WSE-2}~\cite{CEREBRAS} and Groq’s \gls{TSP}\cite{TSP} use coarse-grained \glspl{PE} that implement specialized matrix and vector instructions in conjunction with decoupled data movement instructions to hide memory latencies.

In both approaches, scaling up the shared-L1 cluster within a tile offers clear benefits: better amortization of main memory latency and an improved tile compute vs. boundary bandwidth ratio for operations like matrix multiplication. Ideally, within a tile, every \gls{PE} would access each L1 memory bank directly via an all-to-all flat crossbar interconnect. However, \cite{mempool} demonstrated that achieving physical design feasibility over hundreds of \glspl{PE} requires hierarchical implementations of the in-tile \gls{PE}-memory interconnect. As depicted in Fig.\ref{fig:generic_cluster}, \glspl{PE} and memory banks within a tile are clustered, and connected through high-bandwidth local interconnections for accessing physically neighboring banks. In contrast, accessing physically remote L1 banks requires moving data across higher-latency sections of hierarchical intra-tile interconnect with reduced bandwidth. This limitation arises from the restricted number of access ports at hierarchy boundaries, which are designed to bound the complexity of intra-cluster interconnection routing. This necessary compromise induces \gls{NUMA} effects on L1, leading to stalls and underutilized \glspl{PE}; mitigating these effects is essential to preserving efficiency in scaled-up tiles.

This paper addresses the \gls{PE} utilization bottleneck caused by L1 memory \gls{NUMA} latencies in large \gls{PE} clusters through address runtime remapping. We propose \gls{DAS}, a flexible, adaptable, runtime-configurable address mapping technique. \gls{DAS} remaps contiguous address spaces to physically adjacent memory banks based on the workload's memory access patterns, placing the data physically close to \glspl{PE}. This mechanism is tightly integrated with a software \textit{dynamic data allocator}: a runtime-managed linked list tracks free space within the L1 memory and allocates data according to specified \gls{DAS} configurations. \gls{DAS} enhances data locality and ensures high \gls{PE} utilization, providing both processing efficiency and architectural flexibility.

We implement and validate \gls{DAS} on the open-source~TeraPool cluster\cite{terapool}, the largest shared-L1 multiprocessor tile reported in the literature. Compared to the baseline TeraPool, which uses a fixed, unmodifiable L1 address mapping, our \gls{DAS} approach achieves speedups of up to 3.68$\times$, 2.77$\times$, and 2.20$\times$ on key kernels in transformer training and inference (\gls{GEMV}, \gls{GEMM}, and Self-Attention), with high \gls{PE} utilization (0.76, 0.86, and 0.80, respectively). We further demonstrate the effectiveness of our \gls{DAS} design on real-world model deployment, benchmarking on the \gls{vit}-L/16 model. Each encoder layer executes in \SI{5.67}{\milli\second}, with 0.81 PE utilization and a 1.94$\times$ speedup compared to the fixed word-level interleaved baseline. The key contributions of this paper are:

\begin{itemize}
    \item We propose the \textbf{Dynamic Allocation Scheme (DAS)}, a runtime-configurable address mapping scheme for scaled-up clusters of programmable \glspl{PE} with multi-banked shared-L1 data memory. An address remapping mechanism is introduced to remap contiguous L1 memory addresses to a subset of physically neighboring memory banks, improving data locality by creating a finer memory partition granularity. We extend the L1 heap with a dynamic memory region, where data allocation with \gls{DAS} is enabled. 
    \item We demonstrate \gls{DAS} integration within an ultra-large-scale shared-L1 cluster, featuring \circled{1} an address mapper that implements the remapping logic for \glspl{PE} and the \gls{DMA} engine, \circled{2} a unified dynamic allocator that manages the dynamic heap region, supporting multiple concurrent mapping schemes, and \circled{3} a runtime library with an \gls{API} for seamless integration and usage.
    \item We evaluate \gls{DAS}'s performance using key transformer kernels and \gls{ViT} models, compared to implementations in the unmodifiable address mapping baseline. Additionally, we assess its impact on \gls{PE} utilization and throughput by comparing the results with the NVIDIA A100 GPU.
\end{itemize}
Implemented with GlobalFoundries' \SI{12}{\nano\meter} FinFET technology, our \gls{DAS} design incurs \textless \SI{0.1}{\percent} logic area overhead in TeraPool, without introducing critical timing paths compared to TeraPool baseline. 
Our design is fully open-source\footnote{https://github.com/pulp-platform/mempool}.

% \begin{comment}

\section{Dynamic Allocation Scheme}
This section introduces \gls{DAS}, our runtime-configurable address mapping scheme for scaled-up multi-banked shared-L1 parallel multiprocessors. The solution consists of two parts: a programmable address mapper and a dynamic data allocator.

\subsection{Programmable Address-Remapping}
\label{sec:address_scrambling}

Address remapping techniques are commonly employed in systems featuring fixed-function accelerators\cite{acc_1, acc_2} that are tightly coupled to a shared L1 memory. These techniques rearrange the data layout to achieve full bandwidth utilization by relying on a fixed memory access pattern determined by the accelerator’s function. Consequently, address mapping is configured statically at system initialization. Similarly, \gls{PIM} accelerators\cite{pim_1, pim_2} employ bit-shuffle techniques on bank, channel, and column DRAM addresses to reduce storage overhead and improve memory transaction efficiency.

On the contrary, Many-core clusters of programmable \glspl{PE} can execute a wide range of computations with diverse memory access patterns. Yet, they often struggle to achieve full PE utilization due to the \gls{NUMA} effect implied by scale-up. Clearly, a fixed address mapping in L1 is suboptimal. \gls{DAS} offers the flexibility to adapt the address mapping at runtime based on observed memory access patterns, thereby optimizing memory bandwidth utilization for each specific application.

The L1 of a cluster with shared, multi-banked memory is a continuous address space. A \textit{sequentially interleaved} address mapping scheme is commonly employed as a baseline to minimize bank conflicts. In this scheme, consecutive memory addresses are distributed across the same row in different banks first, before moving to the next row within each bank, and cyclically reaching all the banks in the L1 memory.

\begin{figure}[ht]
  \centering
  \includegraphics[width=\linewidth]{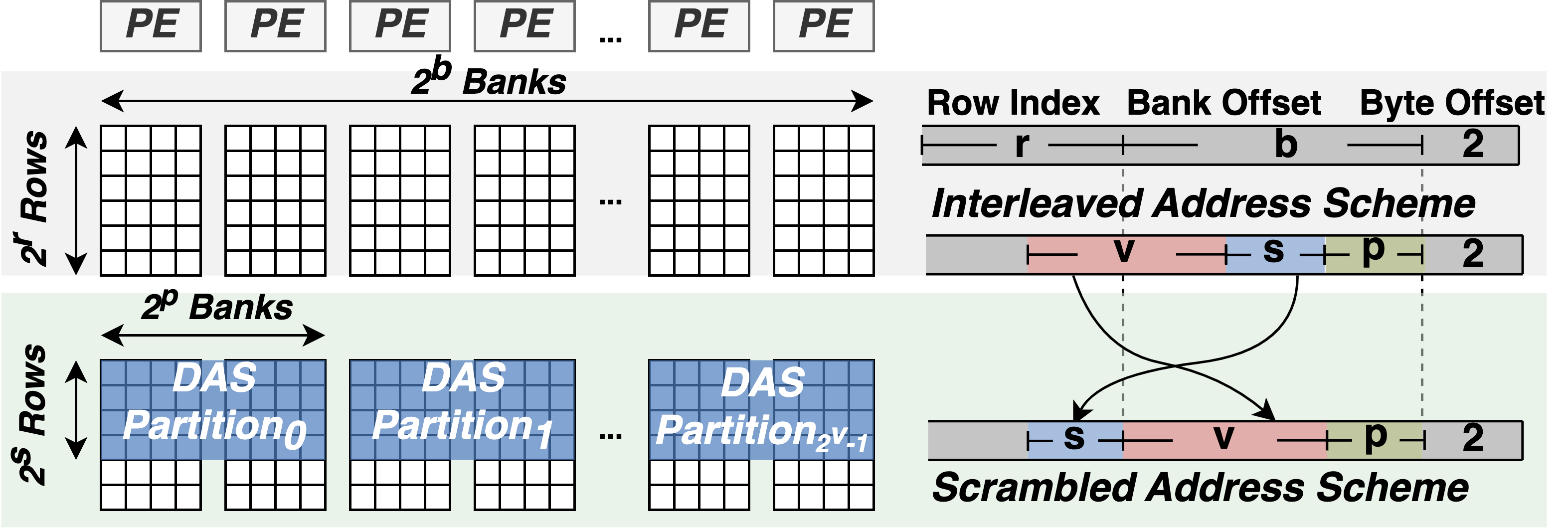}
  \caption{Address-remapping logic for the shared, multi-bank L1 memory. The width of section $p$ represents the \gls{DAS}-partition granularity and the width of section $s$ denotes the size of the partition region.}
  \label{fig:address_scrambling}
    \vspace{-1em}
\end{figure}

Fig.\ref{fig:address_scrambling} describes the interleaved mapping scheme in a cluster with 4 bytes per word $2^b$ memory banks, $2^r$ rows per bank: 2-bits for the byte offset, $b$ bits identify one of the $2^b$ banks, the remaining $r$ bits serve as the row offset within each bank.

Ideally, every memory bank would be accessible via a dedicated link from every PE. This implies an all-to-all crossbar interconnect between PEs and L1 banks, which is not physically implementable beyond a few tens of endpoints, typically between 16 and 32\cite{mempool}. In a physically implementable large-scale shared-L1 cluster\cite{terapool}, \glspl{PE} access physically neighboring memory banks via high-bandwidth, low-latency local interconnections. Conversely, accessing remote L1 banks within the cluster relies on interconnections with higher latency and reduced bandwidth, a limitation caused by the constrained number of \gls{PE}-to-L1 interconnects. This induces \gls{NUMA} effects when accessing L1 memory banks. When applying the interleaved address scheme, more than 99\% of contiguous memory requests target remote L1 banks with higher latency. Ideally, data structures shared among \glspl{PE} should be distributed throughout the cluster to balance access. Consecutive requests from the same \gls{PE} to its privately accessed data structure should be confined to the same group of physically adjacent banks to leverage uniform low-latency local interconnections and reduce data access overhead.

To achieve this, we propose an \textit{address-remapping} mechanism in L1. The byte offset remains consistent. The next $p$ bits designate a \textit{partition} of $2^p$ banks, ensuring that sequential memory addresses are folded across banks within each partition before proceeding to the next row. An additional $s$-bit specifies that $2^s$ rows are affected by this remapping logic. This enables \gls{DAS}-regions with different address mappings across the L1 address space. Finally, the remaining $v=b+s-p$ bits index the newly formed partitions across the cluster. In summary, this scheme modifies the interleaved mapping by shifting $s$ bits from the bank offset to the beginning of the row index, creating a finer memory partition granularity for improved data locality.

The address-remapping can be implemented as a small combinational logic block connected to the \gls{LSU} of \glspl{PE}. In this fashion, the physical addresses after the address mapper are static, while the memory addressing scheme as seen by the \glspl{PE} can be programmed at runtime.

\begin{figure}[ht]
    \vspace{-0.8em}
  \centering
  \includegraphics[width=\linewidth]{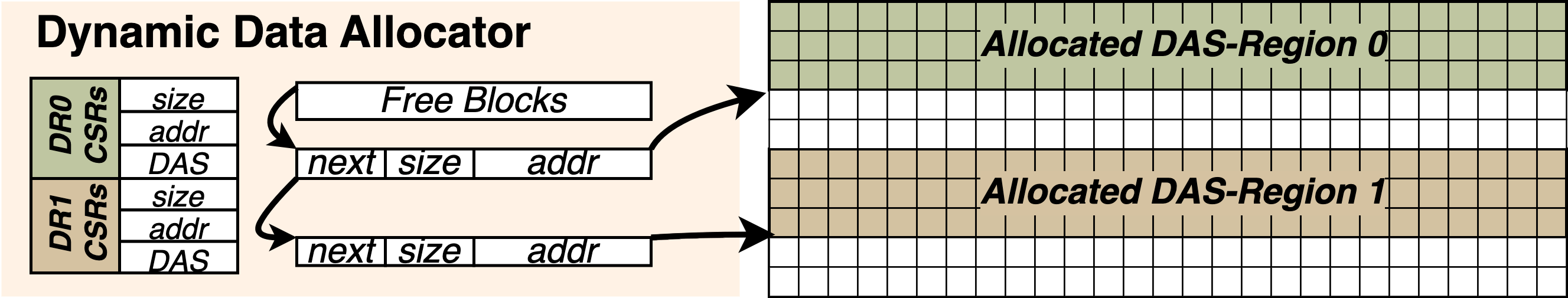}
  \caption{Unified dynamic allocator manages the dynamic heap using a linked list to track free address space and maintains the CSR to configure the address-remapping logic at runtime.}
  \label{fig:allocator}
    \vspace{-1em}
\end{figure}

\subsection{Dynamic Data Allocator}
\label{subsec:allocator}
We restrict the \gls{DAS} address-remapping features to a reserved \textit{dynamic heap region} of L1. Within this region, the data allocation is managed by a unified \textit{dynamic data allocator}, which can keep track of heap \gls{DAS}-regions with multiple concurrent remapping specifications.

Fig.\ref{fig:allocator} illustrates the dynamic data allocator: a runtime-managed software linked list, where each node represents the size and starting address of available address space within the dynamic heap. The nodes are linked in a unidirectional chain, ordered by the starting addresses of their corresponding address space.

At system initialization, the allocator's \textit{free block} pointer is set to the starting address of the dynamic heap region. Upon receiving an allocation request, the allocator traverses the linked list of available address spaces to locate a block large enough to accommodate the requested data. 
Once such a block is found, the allocator updates the list and records the block’s size, starting address, and the programmer-specified \gls{DAS}-region remapping scheme. The starting address of the allocated block is then returned to the program for subsequent data transfer and computation, while the associated \gls{DAS}-region information configures the address-remapping logic. 
Once the data stored in a \gls{DAS} region is no longer required for computation, the corresponding memory block is deallocated by reinserting it into the free-block list and merging it with adjacent free blocks when applicable. The associated \gls{DAS}-region configurations are cleared to allow future allocations.

The multi-region approach aligns memory addresses to the operands' individual access patterns, optimizing data locality within L1 to minimize undesirable \gls{NUMA} effects. 
\section{DAS Integration in Large-Scale Shared-L1 Clusters}
% \section{DAS Integration}

\gls{DAS} is a versatile design suitable for clusters of programmable \glspl{PE} with a shared multi-banked memory of any size. However, banking conflicts in memory accesses to data structures in the same banks grow linearly with the number of \glspl{PE}. Moreover, in big clusters, the hierarchical implementation of \gls{PE}-to-L1 interconnects introduces \gls{NUMA} effects, further impeding computational efficiency.

Therefore, our evaluation of \gls{DAS} focuses on scaled-up shared-L1 many-core clusters. We first assess how \gls{DAS} enhances memory locality and improves performance in systems that push the \gls{SPMD} programming model to its limits, which exacerbates the banking conflicts and \gls{NUMA}-related memory access stalls. Additionally, we find in the large shared-L1 of these systems a rich design space to explore optimal \gls{DAS} allocation patterns and implement in-L1 data processing pipelines.

\begin{figure}[ht]
  \centering
  \includegraphics[width=\linewidth]{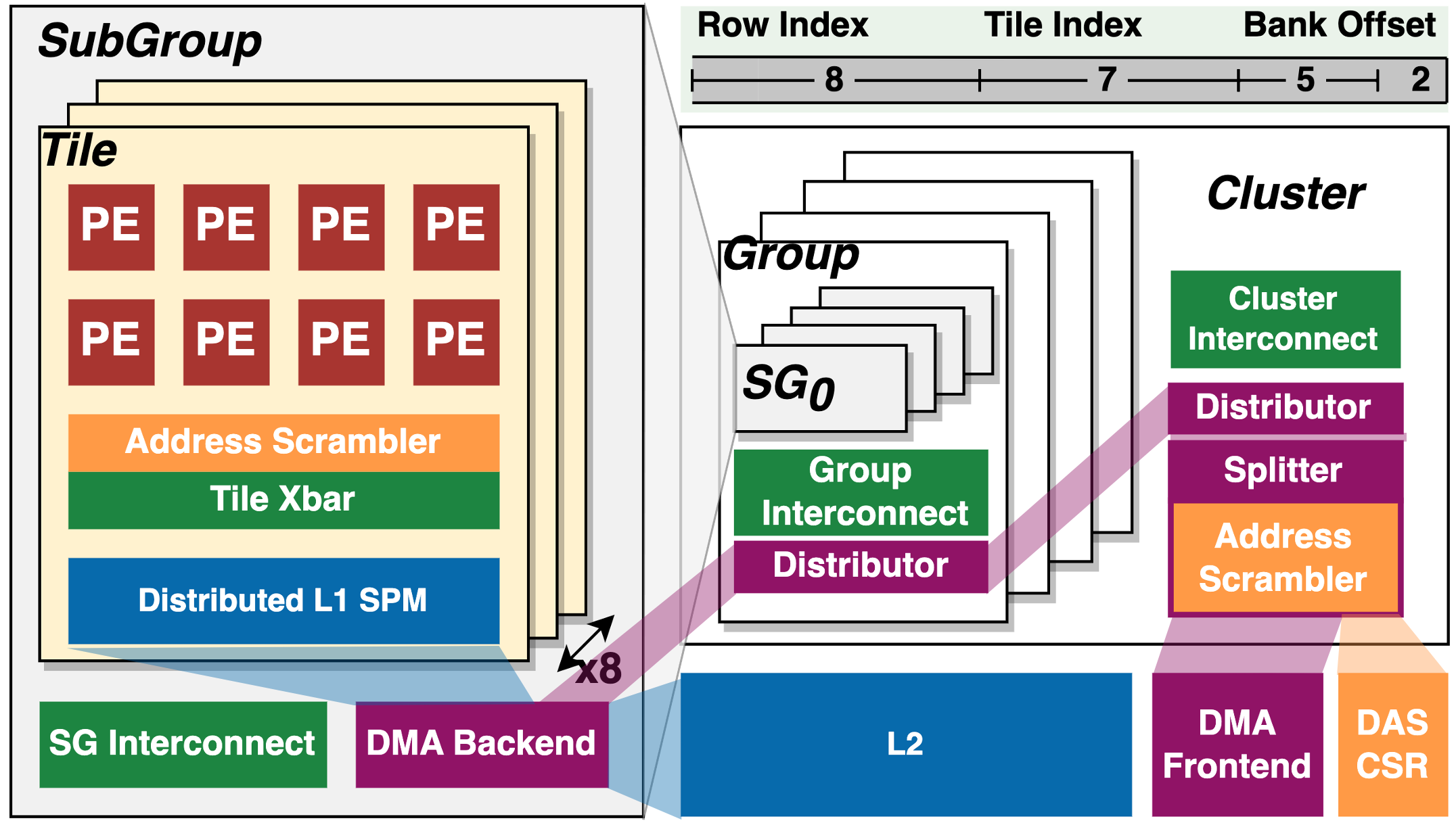}
  \caption{Terapool architecture with DAS integration. Each Tile contains 8 \glspl{PE} connected with 32 memory banks with a fully connected crossbar. Data access on remote memory banks is routed with a hierarchical interconnection marked in green. A modular DMA engine is responsible for data movement from and to L1 memory.}
  \label{fig:TeraPool}
  \vspace{-1em}
\end{figure}

We demonstrate the integration of the proposed \gls{DAS} design within TeraPool, an open-source 1024-\gls{PE} shared-L1 parallel multiprocessor. As the largest implementation of the shared-L1 parallel multiprocessor in literature, TeraPool presents a challenging deployment scenario for \gls{DAS}. This section outlines the key architectural modifications required for \gls{DAS} integration, with the detailed architecture illustrated in Fig.\ref{fig:TeraPool}.

\subsection{The TeraPool Parallel Multiprocessor}
TeraPool is a manycore cluster with 1024 compact RV32IMA \glspl{PE} sharing 4 MiB of 4096-banked L1 \gls{spm}. $8$ \glspl{PE} and $32$ memory banks are grouped into \textit{Tiles}. $8$ Tiles form a \textit{SubGroup}, and $4$ SubGroups make a \textit{Group}. The TeraPool \textit{Cluster} is arranged in a $2\times2$ grid of Groups. Memory within a Tile is accessed in 1 cycle via a fully-connected crossbar. Access to L1 memory in other Tiles relies on a hierarchical interconnection. Tiles within a Subgroup communicate via an $8\times8$ crossbar; they access Tiles in other SubGroups of the same Group via an additional three $8\times8$ crossbars. Tiles of a Group access the memory in Tiles of other Groups via three $32\times32$ crossbars. Interconnect hierarchy implies that the access latency is non-uniform: 3 cycles for local SubGroup access, 5 cycles for local Group access, and 7 cycles for remote Groups without contentions in the shared interconnects.

TeraPool has word-level interleaved addresses across all memory banks. $22$ bits are used to index the 4 MiB L1 SPM address space. The lowest 2 bits are the byte offset, $5$ bits index $32$ memory banks in each Tile, $7$ bits index $128$ Tiles, and the remaining bits define the row offset within the bank.

In TeraPool, the programmer handles explicit \gls{DMA} transfers between the higher-level L2 memory and the L1-\gls{spm}. To support block transfers to the distributed L1, an AXI interconnect and a modular DMA engine, iDMA\cite{idma}, consisting of 3 parts, is designed. Any \gls{PE} can program a single register-based frontend via AXI, determining source-target addresses and transfer size. A iDMA midend control panel orchestrates the transfer. It has two modules: the \textit{splitter} and the \textit{distributor}. The splitter receives a single input DMA request from the frontend registers at a time, and segments it into sequential requests at the address boundary that spans one line of L1 SPM. The distributors assign these segmented requests to a backend per SubGroup, ensuring alignment with the SubGroup memory region. The backends act as data movers and issue memory requests to the system level L2, via AXI, receiving data and distributing it to the Tile \gls{spm} banks.

\subsection{DAS Integration}

The hardware implementation of \gls{DAS} is based on an address mapper module integrated immediately after the PE data request interface. This maps all data accesses within the dynamic heap regions according to the specified \gls{DAS} configuration of each region. The address mapper module is also instantiated in the \gls{DMA}-midend splitter to reshape data transfer requests based on memory partition boundaries. 

The address mapper is runtime configurable via \glspl{csr}. The configuration, managed by the dynamic data allocator, includes the partition granularity ($p$), the size of the \gls{DAS}-region ($s$), and the start address of each allocated region, written via the AXI bus.

% The address scrambler is runtime configurable via \glspl{csr}. The configuration of the partition granularity ($p$), the size of the partition ($s$), and the start address of each allocated region are written to \glspl{csr} via the AXI bus.

The unified allocator, described in Sec.\ref{subsec:allocator} manages the dynamic heap.
% , supporting four regions with distinct \gls{DAS} configurations
An \gls{API} for \gls{DAS} is integrated into the TeraPool C-runtime and is organized into three primary steps: \circled{1} \textit{System Initialization}: The dynamic allocator is initialized using the starting point of the dynamic heap region as specified in the linker script. 
\circled{2} \textit{Data Allocation}: At each new \texttt{malloc} request, the allocator reserves memory regions with specified size and writes the \gls{DAS} configuration in the \gls{csr}. It returns the start address of the allocated space for subsequent data transfers. 
\circled{3} \textit{Data Transfer}: The \gls{DMA}-frontend is configured with transfer size, the source address in L2, and the destination address returned by the allocator. Data is then transferred.
\section{\gls{DAS} Optimized Transformer operators}

\label{sec:benchmark}

Attention-based models rely on a set of fundamental computational kernels, namely \gls{GEMV}, \gls{GEMM}, and \gls{sa}. In this section, we demonstrate how the proposed \gls{DAS} optimizes data allocation for these kernels by tailoring memory partitions to their individual access patterns. We further illustrate the adaptability of \gls{DAS} by applying it to the transformer encoder layer and evaluating its impact on end-to-end \gls{vit} models. Since the ViT architecture incorporates the same core operations as other mainstream attention-based models, our results suggest that the benefits of \gls{DAS} can be generalized across a wide range of transformer architectures.

\begin{figure*}[ht]
    \vspace{-1em}
  \centering
  \includegraphics[width=\linewidth]{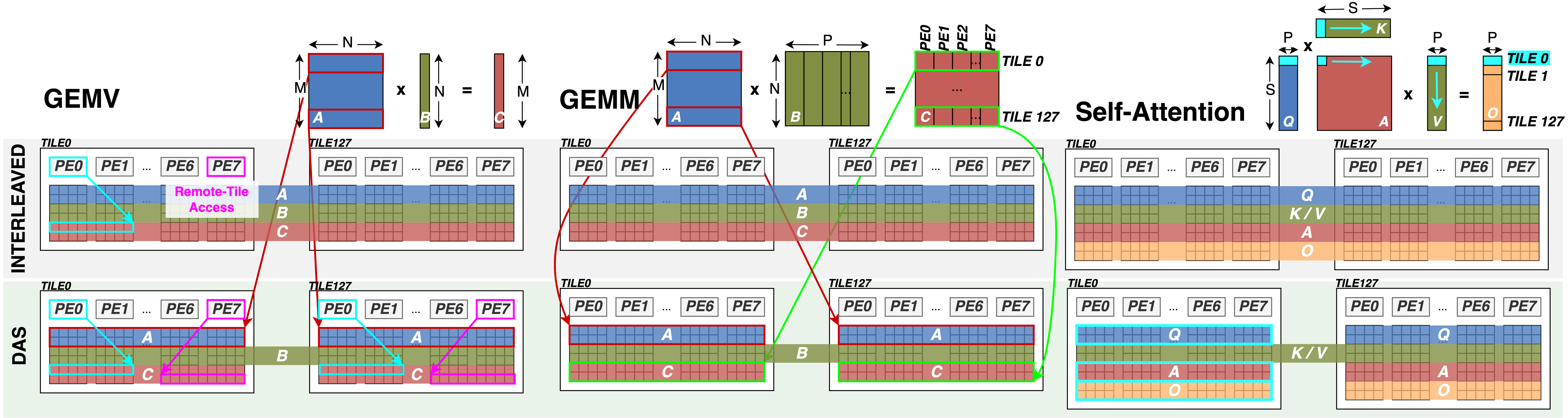}
  \caption{Mapping of \gls{GEMV}, \gls{GEMM}, and Self-Attention with Fully-Interleaved Scheme (upper) and \gls{DAS} (lower).}
  \label{fig:gemm_mapping}
    \vspace{-1em}
\end{figure*}

The maximum kernel size without output tiling is chosen to fully utilize the cluster's 4MiB L1 memory, as this configuration presents the most challenge for mitigating \gls{NUMA} effects. For larger kernels, the double buffering strategy is employed. Fig.~\ref{fig:gemm_mapping} illustrates the memory layout difference between our \gls{DAS} implementation and the fixed addressing scheme baseline.

\subsection{GEMV and GEMM Implementation}
\label{sec: gemm_opt}

\subsubsection{\gls{GEMV}}
In $c=A\times b$, where \( A \in \mathbb{R}^{M \times N} \), \( b \in \mathbb{R}^{N} \), and \( c \in \mathbb{R}^{M} \), outputs are divided between $N_{PE}$ parallel \glspl{PE}. Each \gls{PE} computes $M/N_{PE}$ outputs at contiguous addresses. In the inner loop of our two-loop implementation, blocks of 4 $A$-rows are fetched by \glspl{PE} to produce 4 elements in $c$.

Since output elements are exclusively accessed by individual \glspl{PE}, \gls{DAS} maps the addresses corresponding to $M/N_{PE}$ output elements for each \gls{PE} within a Tile to that Tile’s memory banks, then extends this scheme across the cluster. Exclusively accessed $M/N_{PE}$ rows of the input matrix for each PE in a Tile map to Tile banks. The input vector is accessed by all \glspl{PE}. Therefore, \gls{DAS} maps $b$ fully interleaved to balance access.

When $M<4N_{PE}$, independent \gls{GEMV} problems are assigned to smaller groups of cores, with the described scheme, repeated over the cluster memory. Our evaluation handles computations as large as a single $32\times16384$ \gls{GEMV} computation or 16 $64\times512$ parallel \glspl{GEMV} without output tiling.

\subsubsection{\gls{GEMM}}
$C=A\times B$, where \( A \in \mathbb{R}^{M \times N} \), \( B \in \mathbb{R}^{N \times P} \), and \( C \in \mathbb{R}^{M \times P} \). In the innermost of our 3-loops implementation, a $4\times4$ elements window of $C$ is computed. Blocks of 4 A-rows are fetched by \glspl{PE} of different Tiles, producing blocks of 4 rows in C. The \gls{DAS} maps the addresses of each 4 A and C-rows to banks of different Tiles. $N_{PE/Tile}$ cores in a Tile each fetch from $P/(4\times N_{PE/Tile})$ columns of B to produce as many $4\times4$ C-windows. The B-columns are fetched by \glspl{PE} in all Tiles. Therefore, B is mapped in an interleaved word scheme across the cluster banks. 

When $M/4$ is smaller than the Tile number, independent \glspl{GEMM} are assigned to smaller \gls{PE}-groups. In this case, $B$ addresses are mapped locally to the \gls{PE}-group. Our evaluation scales to handle computations as large as a single $256\times1024\times256$ parallel \gls{GEMM} or 32 independent $64\times64\times64$ \glspl{GEMM} without output tiling.

\subsection{Self-Attention}
\label{subsec: sa}

\begin{figure}[ht]
  \centering
  \includegraphics[width=\linewidth]{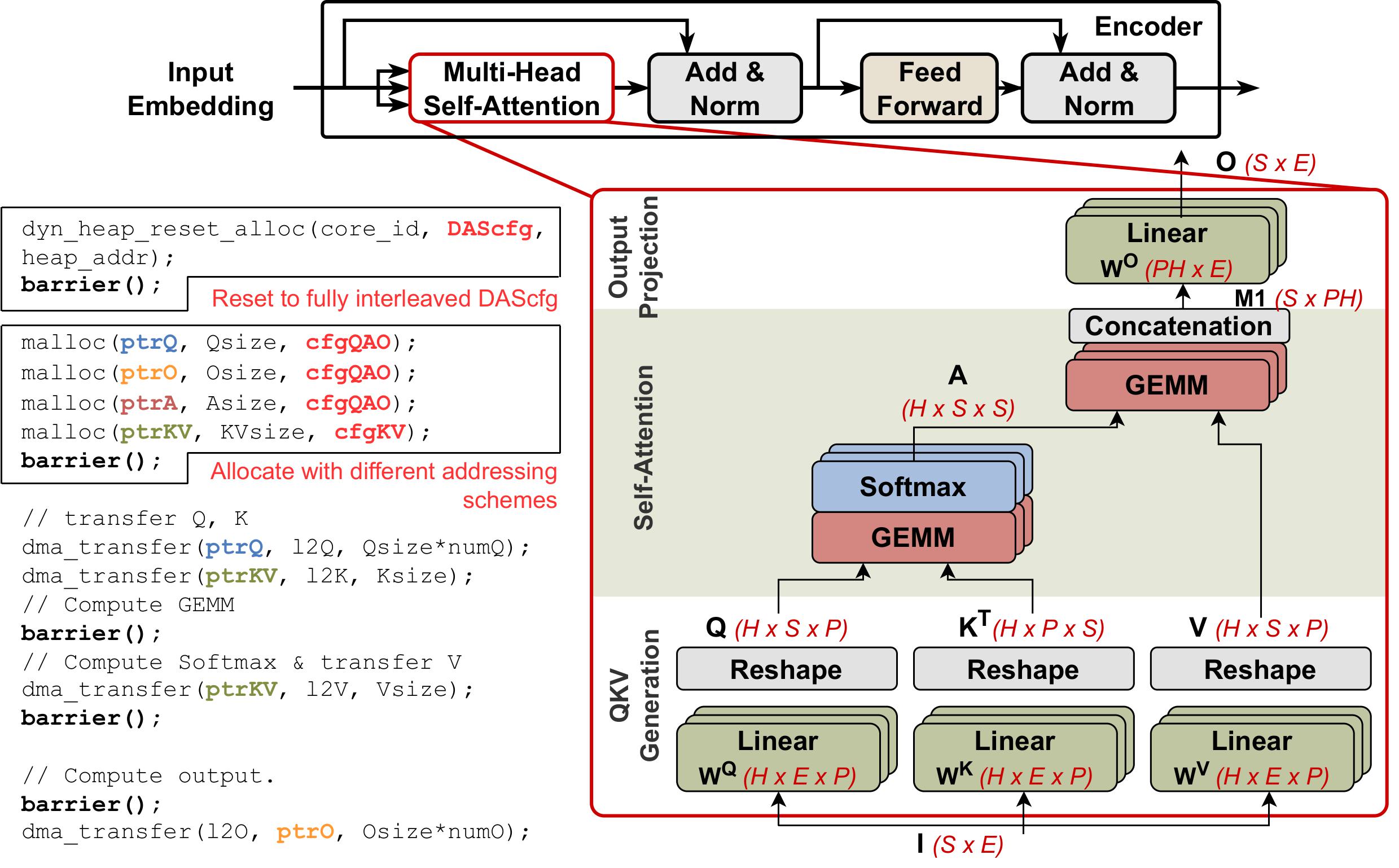}
  \caption{Topology of Transformer Encoder block. Self-Attention is highlighted in the \gls{MHSA} block. The pseudocode on the left shows memory allocationa and transfers for the Self-Attention computation.}
  \label{fig:fa2_flow}
    \vspace{-1em}
\end{figure}

Fig.~\ref{fig:fa2_flow} illustrates the overall architecture of an encoder block and provides a detailed breakdown of the \gls{MHSA}. Given an input sequence \( I \in \mathbb{R}^{S \times E} \), we apply three distinct linear transformations to $I$ for each attention head and generate the query \( Q \in \mathbb{R}^{H \times S \times P} \), key \( K \in \mathbb{R}^{H \times S \times P} \), and value \( V \in \mathbb{R}^{H \times S \times P} \) matrices (with sequence length $S$, number of attention heads $H$, embedding dimension $E$, and head dimension $P$ = $E/H$). Each transformation uses a unique set of weights to project $I$ into a head-specific subspace. 

For each head $i \in (1, H )$, the attention output $O_i$ is computed in three steps: 
% \begin{align*}
% A_{i}^\prime&={Q_iK_i^T}/\sqrt{d} \\
% A_i&=softmax(A_i^\prime) \\
% O_i&=A_i \times V_i
% \end{align*}
\begin{equation}
A_{i}^\prime=\frac{Q_iK_i^T}{\sqrt{d}}, \quad A_i=softmax(A_i^\prime), \quad O_i=A_i \times V_i
\end{equation}

\begin{enumerate}
    \item  \textit{Score computation}. The attention scores $A_{i}^\prime$ are computed by taking the dot product of the query and key matrices, and then scaling by the factor $\sqrt{d}$.
    \item \textit{Weight normalization}. The raw scores are converted into normalized attention weights $A_{i}$ by applying the row-wise softmax function.
    \item \textit{Weighted sum}. The final attention output $O_i$ is obtained by performing a weighted sum of the value matrix using the normalized attention weights $A_{i}$.
\end{enumerate}

We implement self-attention using \gls{FA2}\cite{flashattention}. In the following discussion, we focus on a single attention head and omit the subscript $i$ for clarity. 
The \gls{FA2} parallelization scheme is clarified on the top right of Fig.~\ref{fig:gemm_mapping}: the output matrix $O$ is divided across Tiles along the sequence-length dimension $S$, enabling concurrent processing of $N_{Tile}$ segments. The query matrix $Q$ is divided in $N_{Tile}$ corresponding sub-matrices. Each Tile accesses exclusively a query sub-matrix $Q_{j} \in  \mathbb{R}^{\frac{S}{N_{Tile}} \times P}$, $j \in [0, N_{Tile})$, to compute a unique output sub-matrix $O_{j} \in  \mathbb{R}^{\frac{S}{N_{Tile}} \times P}$. \gls{DAS} maps these sub-matrices to memory banks within each Tile and extends this scheme across the entire cluster, as in Fig.~\ref{fig:gemm_mapping}. The corresponding \gls{API} calls are in the pseudo-code of Fig.~\ref{fig:fa2_flow}: we reset the allocator to a fully-interleaved configuration, we then allocate $Q$ and $O$ to memory regions with the same \gls{DAS} configuration (specified by a single parameter). By folding $O$ and $Q$ along the sequence-length dimension, \gls{DAS} leverages the inherent parallelism of attention mechanisms, providing \glspl{PE} high-bandwidth access to elements in the output and query matrices through local interconnects.

To produce $O_{j}$, the \gls{FA2} tiling strategy is to decompose the large attention score matrix $A_{j}\in  \mathbb{R}^{\frac{S}{N_{Tile}} \times S}$ into $N_{Iter}$ smaller sub-matrices along the sequence-length dimension, denoted as $A'_{j,k} \in  \mathbb{R}^{\frac{S}{N_{Tile}} \times \frac{S}{N_{Iter}}}$, $k \in [0, N_{Iter})$. The key and value matrices are similarly divided into sub-matrices $K_{k}, V_{k} \in \mathbb{R}^{\frac{S}{N_{Iter}} \times P}$. Within each Tile, the \glspl{PE} execute a GEMM on the sub-matrix $Q_{j}$ and the corresponding key sub-matrix $K^T_{k}$, as in Fig.~\ref{fig:gemm_mapping}. This operation yields a partial attention score matrix $A^\prime_{j, k}$. A row-wise softmax is then applied, and the resulting normalized weights $A_{j, k}$ are multiplied with $V_{k}$ to generate a partial output $O_{j, k}$ and a per-token scaling factor. In subsequent iterations, the \glspl{PE} shift to process the next sub-matrices from $K$ and $V$, continuously rescaling and accumulating the final output in $O_{j}$.

The allocation of $A$, $K$, and $V$ is shown in the pseudocode of Fig.~\ref{fig:fa2_flow}. 
During each iteration, the partial attention score matrix $A'_{j, k}$ and its normalized version $A_{j, k}$ are exclusively accessed by each Tile. We map them to memory banks within each Tile, with the same \gls{DAS} configuration used for $Q$ and $O$.
In contrast, sub-matrices $K_{k}$ and $V_{k}$ are shared by all Tiles working on the same attention head. Consequently, \gls{DAS} maps these sub-matrices using an interleaved word scheme across the cluster's memory banks to avoid data duplication and balance memory access. This approach enables the memory-bound softmax operation to fully leverage the high-bandwidth, low-latency local interconnects.

Given the large memory footprint of \gls{FA2}, we devise a transfer strategy to reuse memory locations for $K$ and $V$ matrices. The $Q$ elements are transferred to L1 and remain static for the entire outer loop. For each iteration during the computation of $O_{k}$, once the elements of a sub-matrix in $K_{k}$ are used to produce all the output sub-matrices in $A^\prime_{k}$, they are no longer accessed. Simultaneously, the softmax computation is performed, effectively hiding the transfer of the $V_{k}$ sub-matrix into the memory locations previously occupied by $K_{k}$.

When $S/4$ is smaller than the number of Tiles, independent attention heads are assigned to smaller \gls{PE}-groups. In this scenario, the addresses for $K_i$ and $V_i$ are folded locally within each \gls{PE}-group. Our evaluation scales to process an attention head with 1024 tokens without requiring additional output tiling. For our benchmarks, we evaluate sequence lengths of 1024 tokens by computing the Attention matrix $A_i$ using a tiling strategy with tiles of size $128\times128$, and we consider head dimensions of 32, 64, and 128. 

\subsection{\gls{ViT} Encoder Layer}
\label{subsec:vit} 

\gls{vit} processes 2D image inputs by partitioning the image \( x \in \mathbb{R}^{W \times H} \) into a sequence of flattened 2D patches \( \textbf{$x_p$} \in \mathbb{R}^{S \times T^2 } \). In this formulation, $W$ and $H$ represent the image’s width and height, respectively. Each patch is of size  $T \times T$, and the total number of patches is $S=HW/T^2$. This number corresponds to the effective input sequence length for the \gls{vit} model. Following patch embedding, the \gls{vit} model is constructed as a stack of encoder layers. Each encoder layer is composed of three primary sub-layers: the Normalization layer, the \gls{MHSA} layer, and the \gls{FF} layer, which operate sequentially on the tokenized representations.

The Normalization layer applies token-wise normalization to the input sequence. In our parallel implementation, each \gls{PE} normalizes one token. When the sequence length is smaller than $N_{PE}$ in the cluster, independent Normalization layers are assigned to smaller groups of \glspl{PE}. To minimize bank-access contentions, \gls{DAS} maps addresses of $N_{PE/Tile}$ token vectors to banks in the Tile, and extends the scheme to the cluster.

The \gls{MHSA} layer has three stages: \gls{qkv gen}, \gls{sa}, and \gls{op}. The \gls{qkv gen} stage consists of three \gls{GEMM} operations to compute query, key, and value matrices. Each \gls{GEMM} processes a $(S+1)\times T^2 \times P$ sized matrix, where $S$ is the sequence length, T is the patch dimension, and P is the head dimension. The parallel implementation of these operations is described in Section \ref{sec: gemm_opt}. The input token embeddings remain resident in L1 memory, while the weight matrices and the generated Q, K, and V matrices are transferred between L1 and L2 using the double-buffering strategy to minimize \gls{PE} stalls.

The \gls{sa} stage computes the attention scores, implemented as described in Sec.\ref{subsec: sa}, and processes $ N_{PE}/P $ attention heads simultaneously. In \gls{op} stage, a single \gls{GEMM} operation of size $(S+1)\times P \times T^2$ projects the attended values back to the original embedding dimension, as described in Sec.\ref{sec: gemm_opt}.

\gls{FF} Layer consists of two projections: an upscale projection implemented as a \gls{GEMM} of size $(S+1) \times T^2 \times4T^2$ and a downscale projection of size $(S+1)\times4T^2\times T^2$. Both projections utilize output tiling and double-buffered data transfer to maximize throughput. Additionally, we approximate the GeLU activation function using the $tanh$-based formulation to enhance computational efficiency. \gls{DAS} regions are reconfigured in tandem with data allocation prior to data transactions at each stage.

\section{Results}

We synthesized and performed PnR for our \gls{DAS} design integration in TeraPool using Synopsys Fusion Compiler 2022.03 in GlobalFoundries’ 12LPPLUS FinFET technology. Designed to operate at 920 MHz under nominal conditions (TT/$0.80V$/${25}^\circ C$),  the \gls{DAS}-enhanced TeraPool occupies $68.9,\text{mm}^2$ with a \SI{0.1}{\percent} logic area overhead and no frequency degradation compared to the TeraPool baseline. Benchmark results were simulated in QuestaSim version 2022.3.

% \usepackage{multirow}
% \usepackage{graphicx}
% \usepackage{threeparttable}

\begin{table}[]
\centering
\caption{Summary of kernel performance}
\label{tab:test_performance}
\resizebox{\columnwidth}{!}{%
\begin{threeparttable}
\begin{tabular}{rrrrrr}
\specialrule{1.5pt}{0pt}{0pt}
\multicolumn{1}{c}{\textbf{\begin{tabular}[c]{@{}r@{}}Mapping\\ Scheme\end{tabular}}} &
  \textbf{\begin{tabular}[c]{@{}r@{}}Workload\\ Dimension\end{tabular}} &
  \textbf{\begin{tabular}[c]{@{}r@{}}\#Parallel\\ Workload\end{tabular}} &
  \textbf{\begin{tabular}[c]{@{}r@{}}Utilization\\ (IPC)\end{tabular}} &
  \textbf{\begin{tabular}[c]{@{}r@{}}Throughput\tnote{2}\\ (GFLOPs)\end{tabular}} &
  \textbf{Speedup} \\ \specialrule{1.5pt}{0pt}{0pt}
\multicolumn{6}{c}{\textbf{GEMV (2.5Byte/FLOP)}}                                                                                                                       \\ \specialrule{1.5pt}{0pt}{0pt}
                                                                                          & $64\times512$       & 16 & 0.25 & 174.36  & -                                     \\
                                                                                          & $32\times4096$      & 1  & 0.22 & 160.33  & -                                     \\
\multirow{-3}{*}{\begin{tabular}[c]{@{}r@{}}Fix-addressed\\ Baseline\end{tabular}\tnote{1}}                                                      & $32\times16384$     & 1  & 0.19 & 136.45  & -                                     \\ \hline
                                                                                          & $64\times512$       & 16 & 0.76 & 540.74  & {\color[HTML]{009901} \textbf{3.10x}} \\
                                                                                          & $32\times4096$      & 1  & 0.72 & 520.49  & {\color[HTML]{009901} \textbf{3.25x}} \\
\multirow{-3}{*}{\textbf{This work}\tnote{1}} & $32\times16384$     & 1  & 0.72 & 502.34  & {\color[HTML]{009901} \textbf{3.68x}} \\ \specialrule{1.5pt}{0pt}{0pt}
\multicolumn{6}{c}{\textbf{GEMM (1Byte/FLOP)}}                                                                                                                         \\ \specialrule{1.5pt}{0pt}{0pt}
                                                                                          & $64\times64\times64$     & 32 & 0.31 & 356.57  & -                                     \\
                                                                                          & $256\times256\times256$  & 1  & 0.30 & 357.50  & -                                     \\
                                                                                          & $256\times1024\times256$ & 1  & 0.34 & 412.09  & -                                     \\
\multirow{-4}{*}{\begin{tabular}[c]{@{}r@{}}Fix-addressed\\ Baseline\end{tabular}\tnote{1}}                                                      & $1024\times256\times256$ & 1  & 0.47 & 564.32  & -                                     \\ \hline
                                                                                          & $64\times64\times64$     & 32 & 0.85 & 989.17  & {\color[HTML]{009901} \textbf{2.77x}} \\
                                                                                          & $256\times256\times256$  & 1  & 0.83 & 983.37  & {\color[HTML]{009901} \textbf{2.59x}} \\
                                                                                          & $256\times1024\times256$ & 1  & 0.86 & 1023.70 & {\color[HTML]{009901} \textbf{2.48x}} \\
\multirow{-4}{*}{\textbf{This work}\tnote{1}} & $1024\times256\times256$ & 1  & 0.74 & 883.91  & {\color[HTML]{009901} \textbf{1.57x}} \\ \specialrule{1.5pt}{0pt}{0pt}
\end{tabular}
\begin{tablenotes}
\item[1] The implemented kernels do not include hardware-specific optimizations.
\item[2] Throughput is measured under nominal condition (tt\_freq = \SI{920}{MHz}).
\end{tablenotes}
\end{threeparttable}%
}
\vspace{-1em}
\end{table}

\subsection{Transformer operators benchmarks}
\begin{figure}[t]
  \centering
  \includegraphics[width=\linewidth]{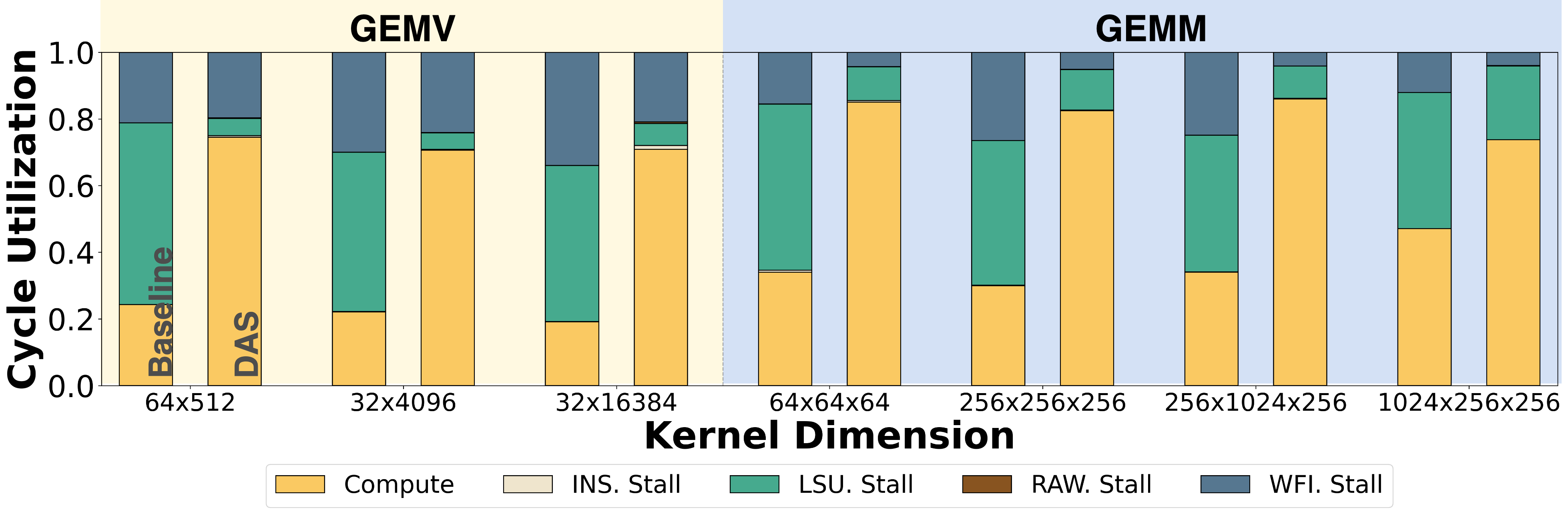}
  \caption{Fraction of instructions and stalls over the total cycles for the key kernels.}
  \label{fig:key-kernel breakdown}
   \vspace{-1em}
\end{figure}

Table \ref{tab:test_performance} summarizes the performance improvements achieved by \gls{DAS} on key kernels. It reports metrics, including \gls{ipc}, throughput, and speedup with respect to the fix-addressed mapping baseline across various kernel dimensions and parallel deployments. These results demonstrate that \gls{DAS} consistently improves computational efficiency across a wide range of arithmetic intensities.

Fig.\ref{fig:key-kernel breakdown} presents the average \gls{ipc} of the parallel implementation of \gls{GEMV} and \gls{GEMM} kernels. We represent a breakdown of idle time due to synchronization (WFI stalls) or architectural stalls: instruction (INS) stalls, load-and-store unit (LSU) stalls, and external pipelined units (RAW) stalls, which originates when the register file of \gls{PE} must wait for the result of long-latency arithmetic instructions.

The primary advantage of \gls{DAS} lies in its ability to sustain high \gls{PE} utilization by significantly reducing \gls{LSU} stalls, achieved through enhanced data locality. As summarized in Table~\ref{tab:test_performance}, our \gls{DAS} implementation achieves up to 3.68$\times$ and 2.77$\times$ speedups over the TeraPool baseline on the GEMV and GEMM kernels, respectively. Higher speedups are observed in kernels with lower arithmetic intensity (i.e., \gls{GEMV}), where a large portion of execution cycles is spent on data access. In these cases, data mapping facilitated by \gls{DAS} mitigates the \gls{NUMA} effects bottleneck, delivering substantial performance improvements.

\begin{figure}[t]
  \centering
  \includegraphics[width=0.9\linewidth]{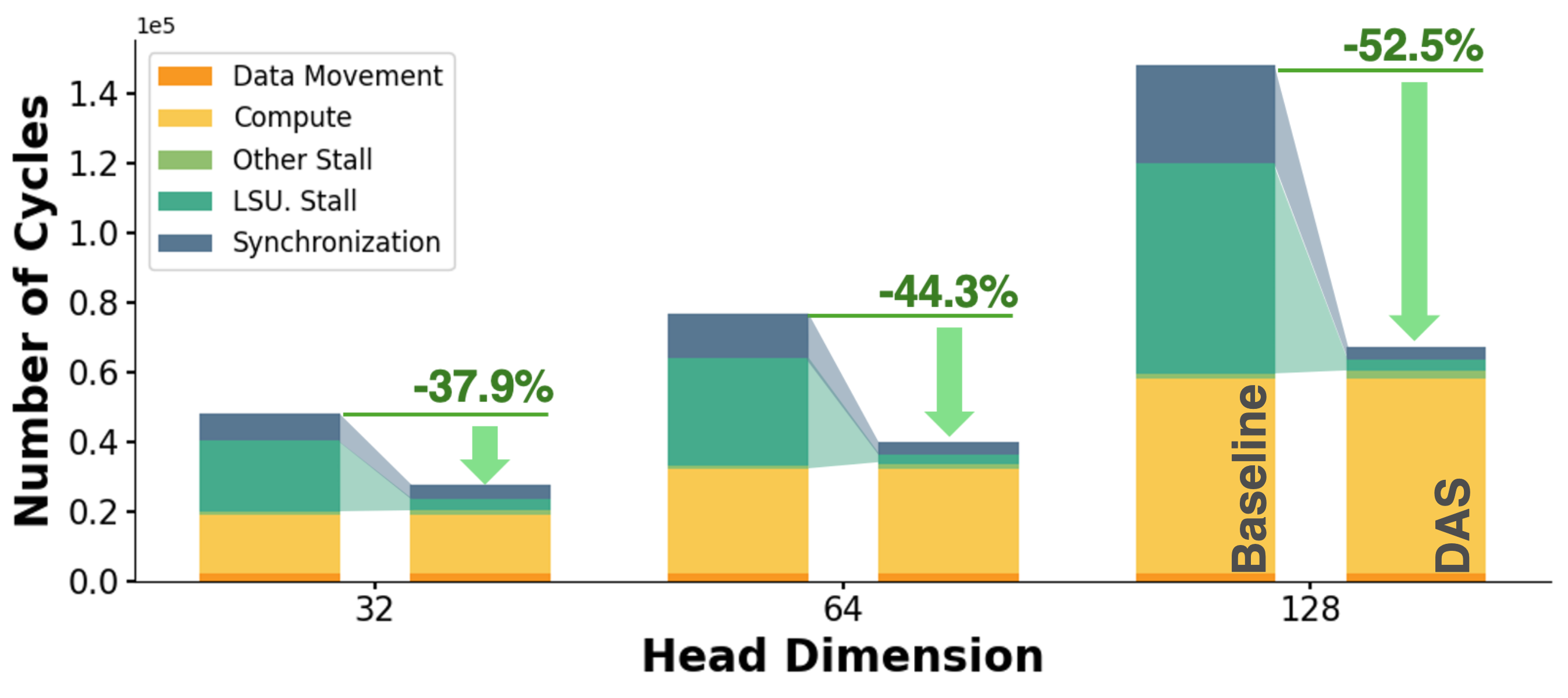}
  \caption{Execution cycle breakdown for the Self-Attention kernel. INS and RAW stalls are combined and labeled as "Other Stall".}
  \label{fig:fa2_breakdown}
  \vspace{-1em}
\end{figure}

Figure~\ref{fig:fa2_breakdown} presents the cycle breakdown of the Self-Attention kernel with a range of hidden dimensions. Cycles used for data movement include those for data allocation, transactions, and the configuration overhead of \gls{DAS} in the enhanced TeraPool. Compared with the fixed-addressed baseline, the DAS-enabled approach requires an equivalent number of cycles for allocation and data transactions while incurring negligible configuration overhead. It reveals a significant reduction in \gls{LSU} stalls thanks to \gls{DAS}-enabled data placement. We highlight that our \gls{DAS} implementation exhibits lower synchronization overhead compared to the fix-addressed baseline, achieved by reducing the scattering of \glspl{PE} during synchronization, as fewer stalls occur during kernel execution. Additionally, asynchronous data movement supported by \gls{DMA} allows memory partition-aware data transfers to be hidden during independent computation phases, further reducing idle cycles and maintaining high throughput. Our \gls{DAS} implementation exhibits increasing speedups as the hidden dimension grows, reaching up to 2.2x in the largest configuration.

\subsection{End-to-end \gls{vit} Encoder Benchmark}

\begin{figure}[t]
  \centering
  \includegraphics[width=0.9\linewidth]{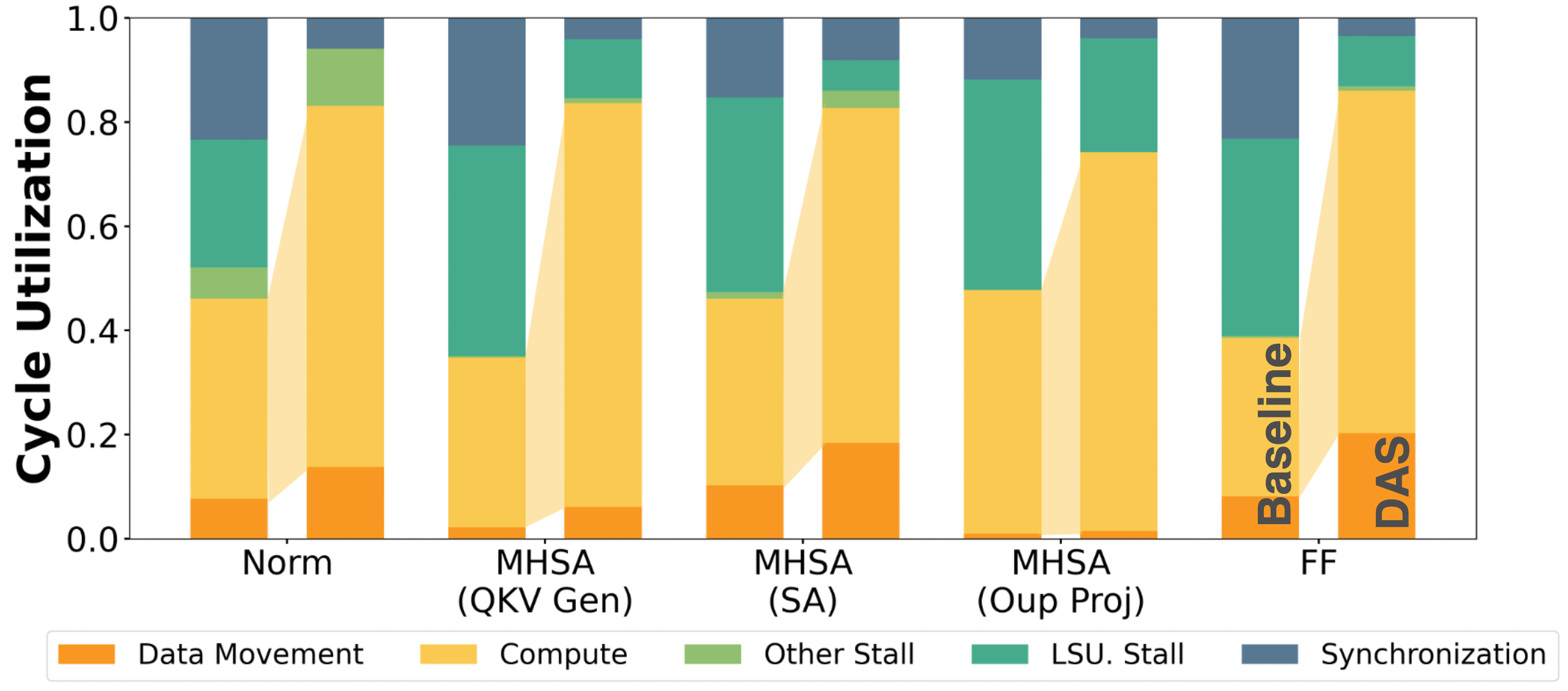}
  \includegraphics[width=0.9\linewidth]{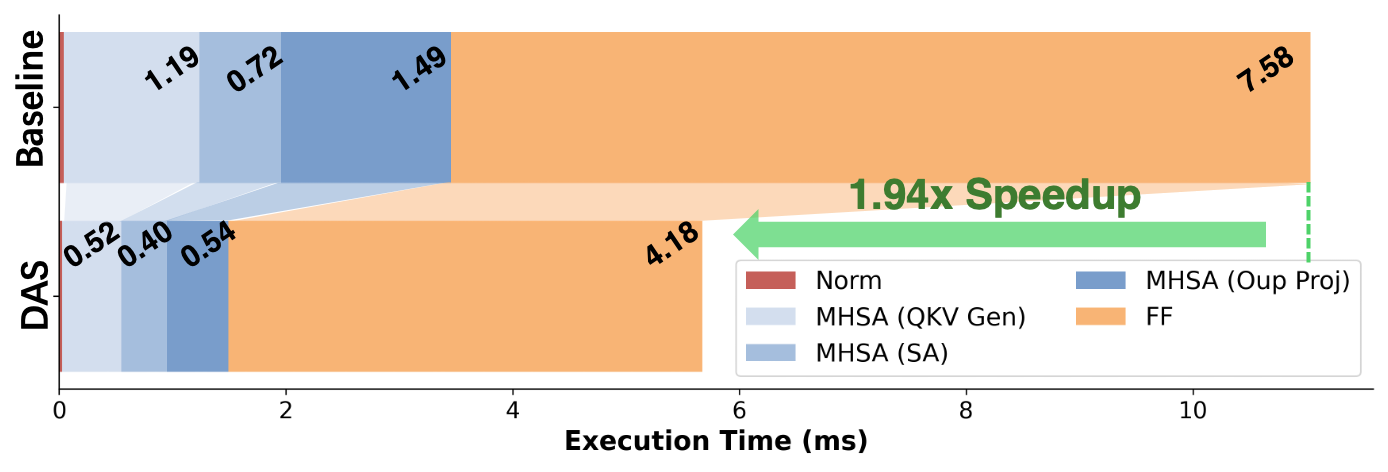}
  \caption{Execution cycle breakdown (upper) and execution time breakdown (lower) for the \gls{vit}-large model.}
  \label{fig:vit_res}
  \vspace{-1em}
\end{figure}

We implemented the encoder layer of a \gls{vit}-L/16 model and evaluated its performance on a layer-by-layer basis. The model processes a $224\times224$ input image by dividing it into $16\times16$ patches, yielding \num{196} tokens plus an additional class token. Each token is represented as a \num{256}-dimensional vector. Fig.\ref{fig:vit_res} presents the fraction of instructions and stalls over the total cycles for each layer. \gls{DAS} consistently facilitates high PE utilization across all layers, with utilization rates reaching 0.80, 0.81, 0.79, 0.83, and 0.79, respectively. Thanks to this high \gls{PE} utilization, each encoder layer is processed within 5.21 million cycles, corresponding to 5.67 milliseconds and 176 images/s throughput at a 920 MHz operating frequency. This performance, achieved on the 12nm FinFET technology node implementation, represents a 1.94x speedup over the baseline TeraPool.

% \usepackage{graphicx}
\begin{table}[]

\centering
\caption{ViT-B/16 Encoder Layer Benchmark Comparison}
\label{tab:vit_compare}
\resizebox{0.9\columnwidth}{!}{%
\begin{threeparttable}

\begin{tabular}{rrrr}
\specialrule{1.5pt}{0pt}{0pt}
\textbf{Platform} & \textbf{A100} & \textbf{TeraPool} & \textbf{This work} \\ \specialrule{1.5pt}{0pt}{0pt}
\textbf{MACs}           & \multicolumn{1}{c}{4.25M} & 4.25M   & 4.25M   \\ \hline
\textbf{\#Cycles}       & 1.45M                     & 18.42M  & 9.46M   \\ \hline
\textbf{\#PEs}          & 6912 \tnote{1}            & 1024    & 1024    \\ \hline
\textbf{\begin{tabular}[c]{@{}r@{}}PE Utilization\\ (IPC)\end{tabular}} & N.A.                      & 0.397 & {\color[HTML]{009901} \textbf{0.809}} \\ \hline
\textbf{\begin{tabular}[c]{@{}r@{}}Throughput\\ (MACs/Cycle)\end{tabular}} & 2923.7 & 230.4 & 448.5 \\ \hline
\textbf{\begin{tabular}[c]{@{}r@{}}Norm. Throughput\\ (MACs/Cycle/PE)\end{tabular}} & 0.423 & 0.225 & {\color[HTML]{009901} \textbf{0.438}} \\ \specialrule{1.5pt}{0pt}{0pt}
\end{tabular}

\begin{tablenotes}
\scriptsize
\item[1] Benchmark runs under FP32 CUDA cores on NVIDIA A100 GPU.
\end{tablenotes}

\end{threeparttable}%
}
\vspace{-1em}
\end{table}

% \begin{tabular}{rrrr}
% \specialrule{1.5pt}{0pt}{0pt}
% \textbf{Platform} & \textbf{A100}             & \textbf{TeraPool} & \textbf{\begin{tabular}[c]{@{}r@{}}Flex-TeraPool\\ (Ours)\end{tabular}} \\ \specialrule{1.5pt}{0pt}{0pt}
% \textbf{MACs}     & \multicolumn{1}{c}{4.25M} & 4.25M             & 4.25M \\ \hline
% \textbf{\#Cycles} & 1.45M                     & 18.42M            & 9.46M \\ \hline
% \textbf{\#PEs}    & 6912\tnote{1}             & 1024              & 1024  \\ \hline
% \textbf{\begin{tabular}[c]{@{}r@{}}Throughput\\ (MACs/Cycle)\end{tabular}} & 2923.7 & 230.4 & 448.5 \\ \hline
% \textbf{\begin{tabular}[c]{@{}r@{}}Utilization\\ (MACs/PE/Cycle)\end{tabular}} & 0.423 & 0.225 & {\color[HTML]{009901} \textbf{0.438}}    \\ \specialrule{1.5pt}{0pt}{0pt}
% \end{tabular}

% \begin{tabular}{rrrr}
% \hline
% \textbf{Platform}     & \textbf{A100} & \textbf{TeraPool} & \textbf{\begin{tabular}[c]{@{}r@{}}Flex-TeraPool\\ (Ours)\end{tabular}} \\ \hline
% \textbf{MACs}         & 4.25M & 2.34M & 2.34M \\ \hline
% \textbf{PEs}          & 6912  & 1024  & 1024 \\ \hline
% \textbf{Cycles}       & 1.45M & 10.14M& 5.21M \\ \hline
% \textbf{\begin{tabular}[c]{@{}r@{}}Utilization\\ (MACs/PE/Cycle)\end{tabular}}  & 0.423& 0.225& {\color[HTML]{009901} \textbf{0.438}}                                   \\ \hline
% \end{tabular}%

We further implemented the \gls{vit}-B/16 encoder layer and summarized the performance comparison in Tab.~\ref{tab:vit_compare} among the baseline TeraPool, our \gls{DAS} implementation, and the NVIDIA A100. The benchmark follows the configuration in \cite{a100-vit}, using a $384 \times 384$ input image with $16 \times 16$ patches. Each encoder layer features an embedding dimension of \num{768} and a head dimension of \num{64}. Performance is measured in terms of throughput (MACs/cycle) and normalized to the number of processing elements (MACs/cycle/\gls{PE}).

To ensure a fair comparison, our evaluation is restricted to FP32 operations. As detailed in \cite{a100-vit}, the A100 GPU is configured to perform FP32 computations using its CUDA cores, with each CUDA core capable of executing one FP32 operation per cycle, matching the performance of a \gls{PE} in TeraPool. This parity allows us to compare the two systems on the same basis without the influence of specialized hardware accelerators.

The results in Table~\ref{tab:vit_compare} indicate that owing to its higher operating frequency and larger number of processing elements, the A100 GPU delivers higher overall throughput. \glspl{PE} in TeraPool are equally capable of executing FP32 operations; however, the \gls{NUMA}-induced reduced bandwidth for remote L1 memory bank access limits their utilization. In contrast, \gls{DAS} enables dynamic address mapping and data allocation that respects each operand's memory access pattern. Exploiting the high-bandwidth local interconnect  effectively improves \gls{PE} utilization without requiring specialized kernel optimization. \gls{DAS} doubles the throughput compared with the fix-addressed baseline, and achieves a normalized throughput of 0.438 MACs/cycle/PE,  surpassing the A100’s 0.423 MACs/cycle/PE. This finding suggests that enabling dynamic address mapping in large-scale shared-L1 clusters can significantly boost \gls{PE} utilization and computational efficiency by tailoring data layouts to the operands’ memory access patterns. Furthermore, it preserves the inherent flexibility and programmability of the manycore cluster.

\section{Conclusions}

This paper introduces the Dynamic Allocation Scheme (DAS), a runtime-configurable address mapping technique that enhances \gls{PE} utilization in large-scale, multi-banked, shared-L1 multiprocessors. By tailoring data layouts to the operands’ memory access patterns, \gls{DAS} achieves a 1.57–3.68$\times$ speedup over the fix-addressed baseline on transformer inference benchmarks, with \textless \SI{0.1}{\percent} increase in logic area.

\section*{Acknowledgment}
\ifx\blind\undefined
    This work is funded in part by the COREnext project supported by the EU Horizon Europe research and innovation program under grant agreement No. \num{101092598}.
\else
    \textit{Acknowledgment information omitted for blind review.}
\fi
% \end{comment}

\bibliographystyle{IEEEtran} % We choose the "plain" reference style
\bibliography{bibliography.bib}

\end{document}